\documentclass[aps,preprint,amsmath,amssymb,nopacs,prb,superscriptaddress,unsortedaddress]{revtex4}

\usepackage{epsf}
\usepackage{graphicx}
\usepackage{sidecap}

\begin{document}

\title{Dirac point spectral weight suppression and surface ``gaps'' in nonmagnetic and magnetic topological insulators}
\author{Su-Yang Xu}\affiliation {Joseph Henry Laboratory, Department of Physics, Princeton University, Princeton, New Jersey 08544, USA}
\author{L. A. Wray}\affiliation {Joseph Henry Laboratory, Department of Physics, Princeton University, Princeton, New Jersey 08544, USA}\affiliation {Advanced Light Source, Lawrence Berkeley National Laboratory, Berkeley, California 94305, USA}
\author{N. Alidoust}\affiliation {Joseph Henry Laboratory, Department of Physics, Princeton University, Princeton, New Jersey 08544, USA}
\author{Y. Xia}\affiliation {Joseph Henry Laboratory, Department of Physics, Princeton University, Princeton, New Jersey 08544, USA}
\author{M. Neupane}\affiliation {Joseph Henry Laboratory, Department of Physics, Princeton University, Princeton, New Jersey 08544, USA}
\author{Chang Liu}\affiliation {Joseph Henry Laboratory, Department of Physics, Princeton University, Princeton, New Jersey 08544, USA}
\author{H.-W. Ji}\affiliation {Department of Chemistry, Princeton University, Princeton, New Jersey 08544, USA}
\author{S. Jia}\affiliation {Department of Chemistry, Princeton University, Princeton, New Jersey 08544, USA}
\author{R. J. Cava}\affiliation {Department of Chemistry, Princeton University, Princeton, New Jersey 08544, USA}
\author{M. Z. Hasan}\affiliation {Joseph Henry Laboratory, Department of Physics, Princeton University, Princeton, New Jersey 08544, USA}

\begin{abstract}

It is predicted that electrons on the surface of a topological insulator can acquire a mass (massive Dirac fermion) by opening up a gap at the Dirac point when time-reversal symmetry is broken via the out-of-plane magnetization. We report photoemission studies on a series of topological insulator materials focusing on the spectral behavior in the vicinity of the Dirac node. Our results show that the spectral intensity is suppressed resulting in a ``gap''-like feature in materials with or without any magnetic impurity or doping. The Zeeman gap in magnetically doped samples, expected to be rather small, is likely masked by the non-magnetic spectral weight suppression involving a large energy scale we report. The photoemission spectral weight suppression observed around the Dirac node thus cannot be taken as the sole evidence for a time-reversal symmetry breaking magnetic gap.  We discuss a few possible extrinsic and kinematic origins of the Dirac point spectral weight suppression (``gap'') observed in the commonly studied topological materials.
\end{abstract}
\date{\today}
\maketitle

Since the discovery of three-dimensional topological insulators (TIs) \cite{RMP, Moore, David Nature BiSb, Kane PRB, Qi PRB}, topological order proximity to ferromagnetism has been of interest. Such interest is strongly motivated by the proposed topological physics and applications including quantized anomalous Hall current, spin current, magnetic monopole image, and inverse spin-galvanic effect \cite{Qi PRB, Yu Science QAH, Axion, Galvanic effect}, all of which are potentially useful for developing novel electronics. The simplest physical scenario used in theoretical proposals \cite{Qi PRB, Yu Science QAH, Axion, Galvanic effect} is that of an energy gap opened at the time-reversal (TR) invariant momenta or the Kramers' point via surface or bulk ferromagnetism. A number of experiments have been performed to study the electronic states near the Dirac point (DP) of TIs with bulk dopants or under magnetic ion deposition on the surface \cite{David Nature 09, Chen Science Fe, Andrew Nature physics Fe, STM Fe, Hofmann, Ast, Valla}. In the surface ion deposition cases, the deposited ions on the surface of a topological insulator is reported to lead to a dramatic reconstruction of the entire surface electronic structure \cite{Andrew Nature physics Fe, Valla, Ast, Hofmann}. It is found to exhibit a rich interplay of Coulomb, magnetic, and disorder perturbations leading to multiple surface states and spectral weight suppression \cite{Andrew Nature physics Fe}. Thus it is not straightforward to isolate the magnetic contribution from that of the nonmagnetic electronic reconstruction components of the surface electronic structure \cite{Andrew Nature physics Fe, Valla, Hofmann}. Whereas in the bulk chemical doping cases, spectral weight suppression at the DP was also reported \cite{Chen Science Fe, Ando QPT} in which the effect is presumed to be of TR breaking origin induced either by introducing bulk magnetic impurities \cite{Chen Science Fe}, or an unusual Higgs mechanism (Higgs bosons) in samples where a topological phase transition is approached by chemical substitution \cite{Ando QPT}. The key experimental finding in these studies is the observation of a strong spectral weight suppression at the DP, leading to a ``twin-peak'' lineshape profile in the photoemission energy distribution curve (EDC), as opposed to a ``single peak'' which represents a robust DP without a gap. In this paper, using angle-resolved photoemission spectroscopy (ARPES), we show that the spectral weight suppression and ``twin-peak'' profile at the DP are observed even on the nominally stoichiometric \cite{stoichiometric} compounds, which does not appear to be related (in a simple way) to factors that are commonly believed to lead to certain symmetry breaking processes, such as the magnetically induced symmetry breaking or some forms of Higgs phenomena. Our studies reported here suggest that the observed spectral weight suppression can arise from extrinsic (nonmagnetic) phenomena as well as due to the special kinematics of the electrons near the singular Dirac node.

% following interpretations of data related to experimental setup and sample surface preparation conditions such as sample cleavage post-relaxation, surface effects due to chemical phase separation during bulk crystal growth, as well as kinematic spatial localization of surface electrons in the vicinity of the DP \cite{Haim Nature physics BiSe, Wray}.

High quality single crystalline samples of topological insulators were grown by the procedure using the Bridgman method \cite{Hor PRB BiSe, Jia, Hasan QPT}. Systematic (10-25 meV) ARPES measurements of the low-energy electronic structure were performed at the Advanced Light Source in the Lawrence Berkeley National Laboratory, the Stanford Synchrotron Radiation Light Source in the Stanford Linear Accelerator Center, and the Synchrotron Radiation Center in Wisconsin. Samples were cleaved {\it in situ} and measured at 10-80 K in a vacuum better than 1 $\times$ 10$^{-10}$ torr, resulting in shiny flat surfaces.

\begin{figure*}
\centering
\includegraphics[width=17cm]{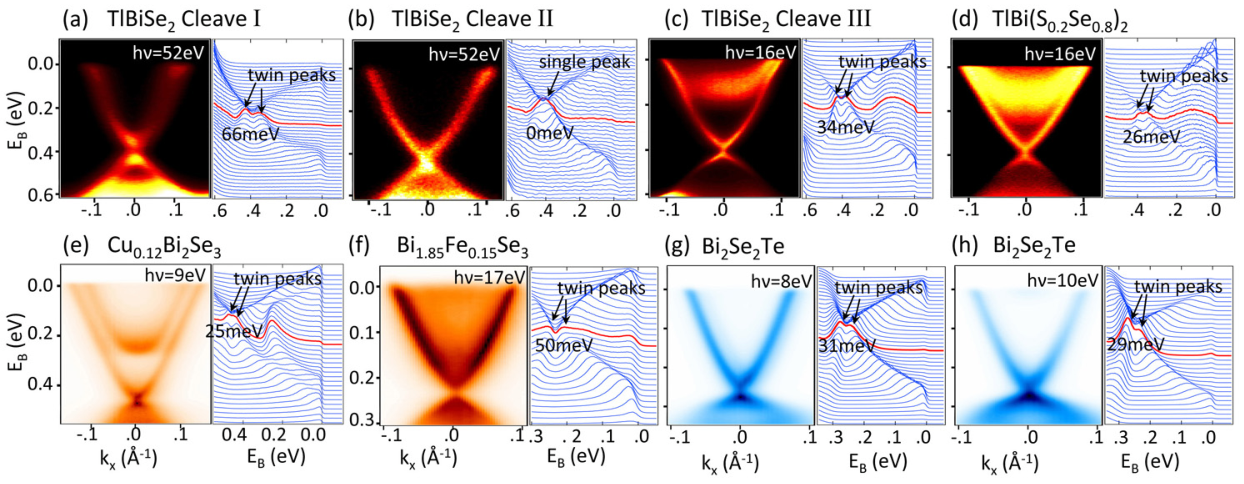}
\caption{\label{Gap1} \textbf{Electronic states in the vicinity of the Dirac point.} High-resolution ARPES dispersion maps and corresponding energy distribution curves (EDCs) in the vicinity of the DP region for different TI compounds. Incident photon energies (h$\nu$) used for the measurements are noted. The TlBiSe$_2$ Cleave I, II, and III (a-c) are from the same growth batch but different pieces of the crystal. The two Bi$_2$Se$_2$Te panels (g) and (f) are from the same cleavage on the same piece of sample using different incident photon energies. The energy scale of the spectral weight suppression at the DP (the energy separation between the ``twin-peak'' in EDC) is obtained by fitting the EDC across DP (highlighted in red) using Lorentzians (as applied in Ref \cite{Chen Science Fe}).}
\end{figure*}

We present high-resolution ARPES dispersion maps on various TI compounds, focusing on the DP regime. Fig.~\ref{Gap1}(a)-(c) show data taken on three different pieces of thallium-bismuth-selenium-sulfur crystals obtained from the same growth batch of the stoichiometric end compound without sulfur substitution, namely TlBiSe$_2$ Cleave I, II, and III. While a clear spectral weight suppression and a ``twin-peak'' behavior are observed for TlBiSe$_2$ Cleave I, Cleave II shows a bright and intact DP featuring ``single peak'' lineshape in the EDC profile. Cleave III also exhibits a ``twin-peak'' profile whose energy splitting between the peaks is relatively weaker than Cleave I. The material with 20\% sulfur substitution TlBi(S$_{0.2}$Se$_{0.8}$)$_2$ [Fig.~\ref{Gap1}(d)] also shows a weak ``twin-peak'' profile. Similar DP spectral weight suppression is observed also in various Bi$_2$Se$_3$ and Bi$_2$Te$_3$ based compounds, as shown in Fig.~\ref{Gap1}(e)-(h). The energy splitting of the twin-peak (the ``gap size'') based on Lorentzian fitting \cite{Chen Science Fe} is listed on each panel. These results suggest that spectral weight suppression observed in classes of TI materials is not uniquely related to magnetic impurities or doping.

\begin{figure*}
\centering
\includegraphics[width=17cm]{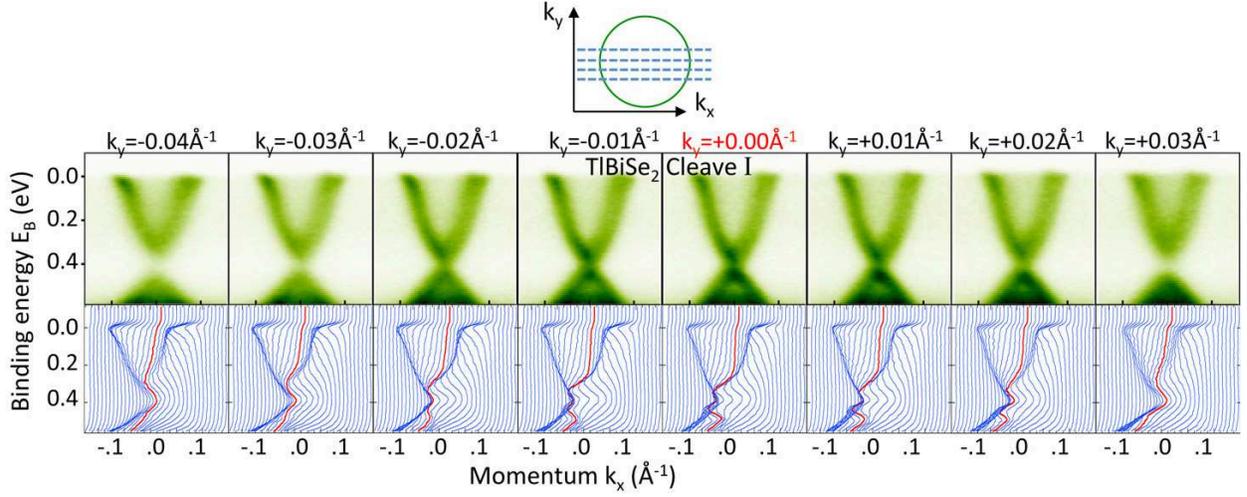}
\caption{\label{Gap2} \textbf{Spectral weight suppression or surface ``gap'' at the Dirac point shown by} $(\textrm{E}_{\textrm{B}}, k_x, k_y)$ \textbf{spectral mapping unrelated to time-reversal breaking (magnetic) effect.} ARPES $(\textrm{E}_{\textrm{B}}, k_x, k_y)$ mapping of TlBiSe$_2$ Cleave I (see Fig.1). A single ARPES dispersion map generates a 2D intensity profile with binding energy ($\textrm{E}_{\textrm{B}}$) versus one direction of the in-plane momentum cut-direction ($k_x$). In order to probe the orthogonal in-plane momentum cut-direction $k_y$, the angle of the sample surface with respect to the electron analyzer slit is finely adjusted.}
\end{figure*}

Systematic studies are performed to rule out possibilities that our observations are due to instrumental errors. First, we show that the DP spectral weight suppression is not due to the photoexcitation energy related matrix element effect caused by the particular choice of the incident energy. Fig.~\ref{Gap1}(g)-(h) show ARPES measurements on a piece of Bi$_2$Se$_2$Te single crystal using identical experimental settings but by varying the incident photon energy. The ``twin-peak'' profile observed in the EDC spectra is found to be independent of the choice of the incident photon energies. In contrast, in the thallium-based compounds, the ``twin-peak" profile is observed in Cleave I, Cleave II shows a ``single peak'' behavior [Fig.~\ref{Gap1}(a) and (b)]. However, Cleave I and II are obtained using identical experimental settings (including same photon energy), but using different pieces of single crystals from the same growth batch. These facts suggest that the DP spectral weight suppression and ``twin-peak'' EDC profile are not due to the choice of certain incident photon energy. We further rule out the possibility that these contrasting observations are due to ``angular misalignment'' in the ARPES scattering geometry (Fig.~\ref{Gap2}). Typically, a single ARPES dispersion map generates a 2D intensity profile as a function of binding energy ($\textrm{E}_{\textrm{B}}$) and one cut-direction of the in-plane momentum ($k_x$). In order to probe the orthogonal in-plane momentum cut-direction $k_y$, the angle of the sample surface with respect to the electron analyzer slit is adjusted accordingly. Therefore, in order to measure the electronic states exactly at the DP ($k_x=k_y=0$), this angle needs to be finely adjusted to lie exactly at normal with respect to the analyzer in order to achieve $k_y=0$ condition. We show a series of ARPES dispersion maps with angular increment of $0.2$ degree (corresponding to the $k_y$ increment of $\simeq0.01$ $\textrm{\AA}^{-1}$ for the specific incident photon energy). It can be seen that the spectral weight suppression persists throughout the $k_y$ mapping procedure, which rules out the possibility of ``angular misalignment''-type instrumental error.

For the nonmagnetically doped compounds or the nominally stoichiometric ones (Fig.~\ref{Gap1}(a)-(e), (g)-(h), and Fig.~\ref{Gap2}), no magnetic ordering or net magnetization response is observed in the bulk susceptibility characterizations, and thus the DP spectral weight suppression observed here cannot be interpreted to be of TR breaking origin. Additionally, in the thallium-based material system, an unusual time-reversal breaking gap opening mechanism is reported in a previous study \cite{Ando QPT}. The thallium-based system is reported to have an intact DP in the absence of a gap in the end compound TlBiSe$_2$ without sulfur substituting selenium, whereas, while finite amount of sulfur is introduced into the system, a gap at the DP is reported to develop due to spontaneous symmetry breaking via a Higgs mechanism. In the same thallium-based system we study here, clear spectral weight suppression is observed even on the end product TlBiSe$_2$ without any sulfur substitution, namely the TlBiSe$_2$ Cleave I and III as shown in Fig.~\ref{Gap1}(a) and (c). Therefore, our systematic results do not support a definite correlation between the DP ``gap'' and the sulfur concentration as described by the unusual Higgs mechanism in the previous study \cite{Ando QPT}. In fact, experience with these materials suggests that for nonmagnetic TI crystals, the DP spectral weight suppression is not systematically reproducible [e.g. see Fig.~\ref{Gap1}(a)-(c)].

We discuss a few possible extrinsic or kinematic origins for the observed Dirac point spectral weight suppression in the materials studied. One such scenario is that of the top-layers-relaxation. It is known that the van der Waals bonding between the quintuple layers (QLs) of Bi$_2$Se$_3$ (or the corresponding unit cell for other layered-structure TIs) is weak, which leads to top-layers-relaxation, especially when disorder and impurities are present on the cleavage plane \cite{Ag, vdW}. The top-layers-relaxation is found to modify the surface states' charge density distribution in the $\hat{z}$ direction \cite{Ag, vdW}. In particular, the charge density distribution of the surface electronic states in the vicinity of the DP is predicted to be pushed down deeper inside the crystal \cite{vdW}. This causes a strong spectral weight suppression at the DP region in photoemission experiments (surface sensitivity less than $8$ $\textrm{\AA}$). The top-layers-relaxation is naturally caused by the mechanical cleavage or exfoliation implied for the surface preparation for ARPES and STM experiments. A surface phase separation due to bulk chemical doping or intergrowth in TI crystals \cite{Ag, Cava Fe} can lead to a similar spectral behavior scenario.

\begin{figure*}
%\centering

\includegraphics[width=17cm]{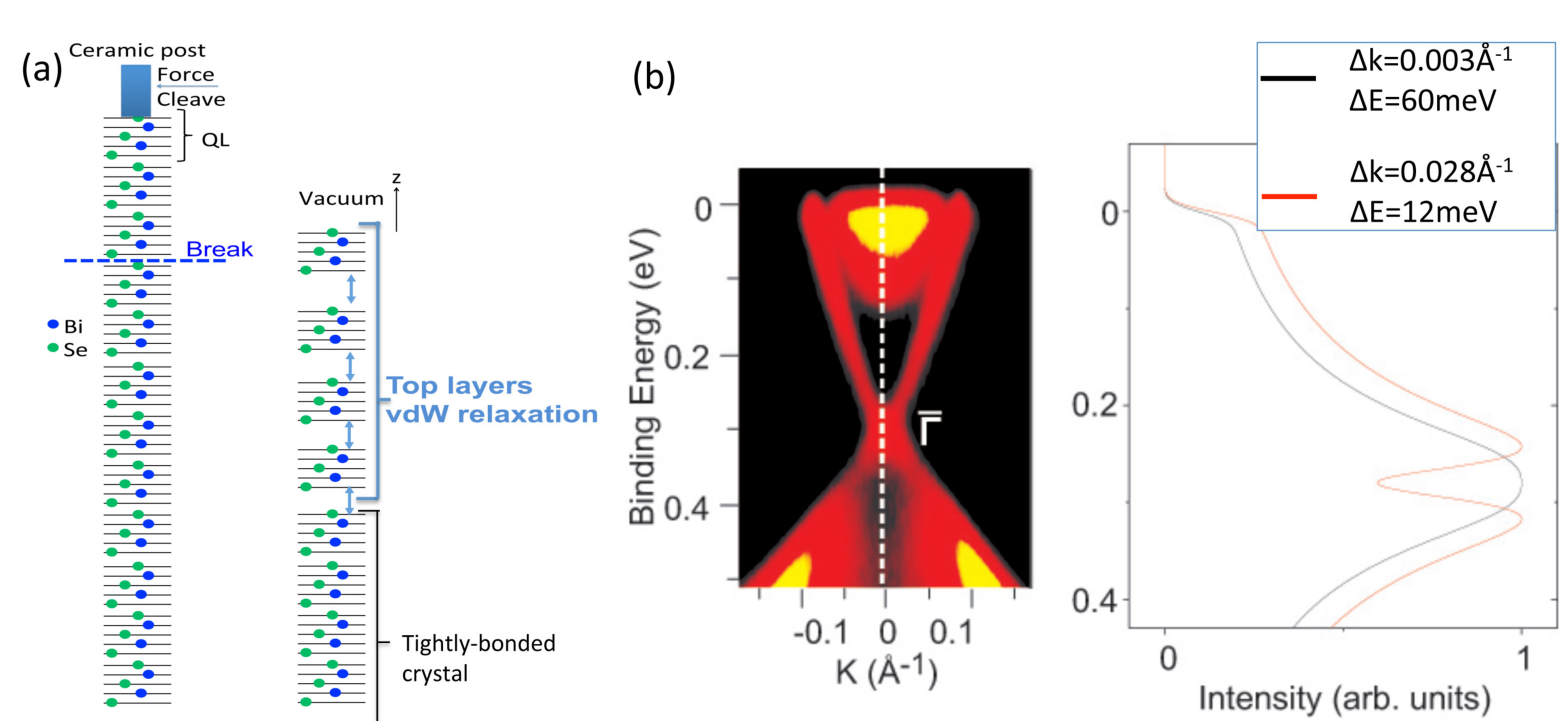}

\caption{\label{Gap3} \textbf{Surface ``gap'' at the Dirac point due to extrinsic or kinematic effects.} (a) Mechanical cleavage leads to finite top-layers-relaxation in samples with weak van der Waals coupling. (b) Variant kinematics near a Dirac singularity. Simulation results show that substantial momentum broadening naturally leads to the ``twin-peak'' lineshape of the density of states along the energy axis in the vicinity of a Dirac singularity. We have chosen two sets of momentum and energy broadening parameters (Lorentzian ${\Delta}k$=0.0275 $\textrm{\AA}^{-1}$, 0.003 $\textrm{\AA}^{-1}$  and ${\Delta}\textrm{E}$=12 meV, 60 meV) representing cases in which either momentum broadening dominates or energy broadening dominates, respectively. The fitted lineshape is then obtained by broadening an ideal Dirac cone with velocity $v=2.3$ $\textrm{eV}{\cdot}\textrm{\AA}$ through Lorentzian convolution in momentum and energy (convoluting each point [$k_{x0}$, $k_{y0}$, $\textrm{E}_0$] with $\textrm{I}\propto\frac{1}{(k_x-k_{x0})^2+(k_y-k_{y0})^2+({\Delta}_k/2)^2}\times\frac{1}{(\textrm{E}-\textrm{E}_0)^2+({\Delta}_\textrm{E}/2)^2}$). With substantial momentum broadening (Lorentzian ${\Delta}k$=0.0275 $\textrm{\AA}^{-1}$, ${\Delta}\textrm{E}$=12 meV), the simulated EDC profile is found to develop a spectral weight suppression at the DP.}
\end{figure*}

The special nature of the electronic structure near the Dirac point (the momentum of the electrons in the vicinity of the Dirac point is very small) also plays an important role in this issue. In reality, the sources of disorder including chemical dopants, impurities and native defects are unavoidable on cleaved surfaces. A recent STM work has shown that \cite{Haim Nature physics BiSe} the energy and momentum of the surface states of TIs are observed to spatially fluctuate in the presence of weak and strong disorders on the surface. For electrons with large momenta (large $k$), the spatial fluctuation of the momentum of the surface electrons ${\Delta}k$ is much smaller than $k$, and hence it does not affect the electronic structure. In contrast, for electrons in the vicinity of the Dirac point ($k\simeq0$), the spatial fluctuation plays a significant role. Since ${\Delta}k{\sim}k$ near the Dirac point, the momentum is not a well-defined quantum number in the sense of diffusive transport. Such a spatial fluctuation suppresses the spectral weight intensity in the vicinity of the Dirac point, resulting in a ``gap''-like feature in spatially averaged measurements \cite{Haim Nature physics BiSe, Mn} such as ARPES. This effect can also be understood by considering the electron kinematics in momentum space. The effect of broadening of electronic states along momentum and energy axes needs to be treated in a special way at Dirac singularities. The finite lifetime of electronic states due to scattering leads to a nonzero energy width. Similarly, the states are broadened along the momentum axes as a consequence of scattering induced localization. DP singularities are a special case in which energy and momentum broadening cause distinctly recognizable effects \cite{Wray}. It is known that the momentum-integrated electronic density of states goes to zero at a DP in the absence of self energy broadening. Momentum broadening does not affect the distribution of electronic states along the energy axis, and thus allows the density of states at the DP to remain zero. Energy-axis broadening (self energy) causes the momentum-integrated density of states at the DP energy to become finite and approach the DP energy in a parabolic dispersion. Accounting for this effect at the DP requires the intrinsic momentum and energy widths to be treated separately. As an example of how such a fit can be carried out, we have chosen two sets of momentum and energy broadening parameters (Lorentzian ${\Delta}k$=0.0275 $\textrm{\AA}^{-1}$, 0.003 $\textrm{\AA}^{-1}$  and ${\Delta}\textrm{E}$=12 meV, 60 meV) representing cases in which either momentum broadening dominates or energy broadening dominates, respectively. The fitted lineshape is then obtained by broadening an ideal Dirac cone with velocity $v=2.3$ $\textrm{eV}{\cdot}\textrm{\AA}$ through Lorentzian convolution in momentum and energy (convoluting each point [$k_{x0}$, $k_{y0}$, $\textrm{E}_0$] with $\textrm{I}\propto\frac{1}{(k_x-k_{x0})^2+(k_y-k_{y0})^2+({\Delta}_k/2)^2}\times\frac{1}{(\textrm{E}-\textrm{E}_0)^2+({\Delta}_\textrm{E}/2)^2}$). As shown in Fig.~\ref{Gap3}(b), with substantial momentum broadening (Lorentzian ${\Delta}k$=0.0275 $\textrm{\AA}^{-1}$, ${\Delta}\textrm{E}$=12 meV), the simulated EDC profile is found to develop a spectral weight suppression at the DP. In addition, it is important to note that both effects we have discussed above are due to non-ideal surface quality and condition of the samples, which is consistent with the fact that the ARPES observed DP spectral weight suppression is surface quality and preparation dependent.

In conclusion, we have provided evidence that a ``gap''-like feature (spectral weight suppression) at the Dirac point observed in various topological insulator compounds with or without magnetic dopants can arise due to nonmagnetic/extrinsic effects which often have energy scales comparable or larger than the magnetic Zeeman gap (theoretically expected in magnetic samples). Therefore, based on the ARPES data alone in the absence of surface magnetization response data, the observation of spectral weight suppression "gap" cannot be taken as the signature of magnetic gap (or exotic symmetry breaking such as the Higgs phenomena in the absence of robust signatures of spontaneous symmetry breaking) on the surface of a topological insulator.

\newpage

\newpage


\begin{thebibliography}{99}
\bibitem{stoichiometric} Note that the ``stoichiometric'' here means no explicit doping during the sample growth. Nevertheless, native defects, e.g. selenium vacancies in Bi$_2$Se$_3$, are generally unavoidable, which causes deviation from the ideal stoichiometry in realistic crystals.
\end{thebibliography}
\end{document}